\documentclass[a4paper,english,reprint, aps]{revtex4}
\usepackage[T1]{fontenc}
\usepackage[latin9]{inputenc}
\usepackage[active]{srcltx}
\usepackage{textcomp}
\usepackage{amsmath}
\usepackage{amssymb}
\usepackage{esint}

\makeatletter

\@ifundefined{textcolor}{}
{%
 \definecolor{BLACK}{gray}{0}
 \definecolor{WHITE}{gray}{1}
 \definecolor{RED}{rgb}{1,0,0}
 \definecolor{GREEN}{rgb}{0,1,0}
 \definecolor{BLUE}{rgb}{0,0,1}
 \definecolor{CYAN}{cmyk}{1,0,0,0}
 \definecolor{MAGENTA}{cmyk}{0,1,0,0}
 \definecolor{YELLOW}{cmyk}{0,0,1,0}
}

\makeatother

\usepackage{babel}
\begin{document}

\preprint{APS/123-QED}

\title{Variational Approach and Deformed Derivatives}

\thanks{The authors wish to express their gratitude to FAPERJ-Rio de Janeiro
and CNPq-Brazil for the partial financial support. We thank Bruno
G. Costa and Ernesto P. Borges for very valuable discussions.}

\author{J. Weberszpil}

\email{josewebe@gmail.com}

\selectlanguage{english}%

\affiliation{Universidade Federal Rural do Rio de Janeiro, UFRRJ-IM/PPGMMC\\
 Av. Governador Roberto Silveira s/n- Nova Igua\c{c}\'u, Rio de Janeiro,
Brasil, 695014.}

\author{J. A. Helay\"el-Neto}

\email{helayel@cbpf.br}

\selectlanguage{english}%

\address{Centro Brasileiro de Pesquisas F\'isicas-CBPF-Rua Dr Xavier Sigaud
150,\\
 22290-180, Rio de Janeiro RJ Brasil. }

\date{\today}
\begin{abstract}
Recently, we have demonstrated that there exists a possible relationship
between $q-$deformed algebras in two different contexts of Statistical
Mechanics, namely, the Tsallis' framework and the Kaniadakis' scenario,
with a local form of fractional-derivative operators for fractal media,
the so-called Hausdorff derivatives, mapped into a continuous medium
with a fractal measure. Here, in this paper, we present an extension
of the traditional calculus of variations for systems containing deformed-derivatives
embedded into the Lagrangian and the Lagrangian densities for classical
and field systems. The results extend the classical Euler-Lagrange
equations and the Hamiltonian formalism. The resulting dynamical equations
seem to be compatible with those found in the literature, specially
with mass-dependent and with nonlinear equations for systems in classical
and quantum mechanics. Examples are presented to illustrate applications
of the formulation. Also, the conserved Nether current, are worked
out.

------
\end{abstract}

\keywords{Deformed Derivatives, Metric Derivatives, Fractal Continuum, Generalized
Statistical Mechanics, Variational Principle, Position-Dependent Mass
}

\maketitle

\section{Introduction}

The minimum action principle implies that, by minimizing some action
variables or functionals, we can obtain the dynamical equations that
describe physical phenomena. The formalism related is known as the
calculus of variations. But, the classical variational calculus has
a major difficult in dealing with nonconservative systems.

Here, we claim that the calculus of variations with deformed-derivatives
\citep{Nosso On conection2015} embedded into a Lagrangian or a Lagrangian
density is adequate to study both, conservative and nonconservative
classical and field systems in order to obtain a version of the respective
Euler-Lagrange equations ($E-L$), combining both cases. Also, systems
with position-dependent mass and nonlinearities can be studied and
the $E-L$ obtained.

The applications of the formalism presented here may include classical
and quantum mechanics, field theory, complex systems and so on. We
are not talking about including classical definitions of fractional
derivatives nor including operators of integer order acting on a d-dimensional
space but we consider a mapping from a fractal coarse-grained (fractal
porous) space, which is essentially discontinuous in the embedding
Euclidean space \citep{Balankin PRE85-Map-2012,Balankin Rapid Comm}
to a continuous one. Also, in the construction of the actions to obtain
the Euler Lagrange equations (ELE), we have used different definitions
for the actions and for the deformed-derivatives and integrals.

In Ref. \citep{Sympletic}, we adopt the viewpoint to suitably treat
nonconservative systems is through Fractional Calculus (FC), since
it can be shown that, for example, a friction force has its form stemming
from a Lagrangian that contains a term proportional to the fractional
derivative, which may be a derivative of any non-integer order \citep{RW}.
Parallel to the standard FC, there is some kind of the local fractional
calculus with certain definitions called local fractional derivatives,
for example, the works of Refs.\citep{Chen-Time Fabic,Kolwankar1,Kolwankar 2,Kolwankar 3,calculus of local fractional derivatives,On the local fractional derivative,Carpintieri}.
Here we are interested in the related approaches with Hausdorff derivative
\citep{Balankin PRE85-Map-2012,Chen-Time Fabic,Hausdorff or fractal derivative}
and the conformable derivative \citep{new definition}. We think that
the most appropriate name of those formalism are deformed-derivatives
or even metric or topological derivatives. All of the mentioned approaches
seem to be applicable to power-law phenomena. Also, for description
of complex systems, the q-calculus, in a non-extensive statistic context,
has its formal development based on the definition of deformed expressions
for the logarithm and exponential \citep{Borges 2004}, namely, the
$q-$logarithm and the $q-$exponential. In this context, an interesting
algebra emerges and the formalism of a deformed derivative opened
new possibilities for, besides the thermodynamical, other treatment
of complex systems, specially those with fractal or multifractal metrics
and presenting long-range dynamical interactions. The deformation
parameter or entropic index, $q$, occupying an important place in
the description of those complex systems, describes deviations from
standard Lie symmetries and the formalism aimed to accommodate scale
invariance in a system with multifractal properties to the thermodynamic
formalism. For $q\rightarrow1$, the formalism reverts to the standard
one.

Here, we consider the relevant space-time/ phase space as fractal
or multifractal\citep{Tsallis-BJF 1999}.

The use of deformed-operators is also justified here based on our
proposition that there exists an intimate relationship between dissipation,
coarse-grained media and the some limit scale of energy for the interactions.
Since we are dealing with open systems, as commented in Ref. \citep{Nosso On conection2015}
, the particles are indeed dressed particles or quasi-particles that
exchange energy with other particles and the environment. Depending
on the energy scale an interaction may change the geometry of space\textendash time,
disturbing it at the level of its topology. A system composed by particles
and the surrounding environment may be considered nonconservative
due to the possible energy exchange. This energy exchange may be the
responsible for the resulting non-integer dimension of space\textendash time,
giving rise then to a coarse-grained medium. This is quite reasonable
since, even standard field theory, comes across a granularity in the
limit of Planck scale. So, some effective limit may also exist in
such a way that it should be necessary to consider a coarse-grained
space\textendash time for the description of the dynamics for the
system, in this scale. Also, another perspective that may be proposed
is the previous existence of a nonstandard geometry, e.g., near a
cosmological black hole or even in the space nearby a pair creation,
that imposes a coarse-grained view to the dynamics of the open system.
Here, we argue that deformed-derivatives, similarly to the FC, allows
us to describe and emulate this kind of dynamics without explicit
many-body, dissipation or geometrical terms in the dynamical governing
equations. In some way, the formalism proposed here may yield an effective
theory, with some statistical average without imposing any specific
nonstandard statistics. So, deformed derivatives and/or FC may be
the tools that could describe, in a softer way, connections between
coarse-grained medium and dissipation at a certain energy scale.

One relevant applicability of our formalism may concern position-dependent
systems (see Ref. \citep{Habib-Revisiting} and references therein)
that seems to be more adequate to describe the dynamics of many real
complex systems where there could exist long-rage interactions, long-time
memories, anisotropy, certain symmetry breakdown, nonlinear media,
etc\citep{Monteiro-Nobre-2013}.  

In the case of quantum mechanics, certain minimum length scale yields
a modification in the position momentum commutation relationship or
some modification in the underlying space-time that may results in
a Schr\"odinger equation with a position-dependent mass \citep{Costa-Almeida-}.

Also, to find more suitable ways to explaining several complex behaviors
in nature, Nonlinear (NL) equations have become an important subject
subject of study \citep{NRT-2012}, since the applicability of linear
equations in physics is usually restricted to idealized systems \citep{NRT-PRL-2011}.

An important point to emphasize is that the paradigm we adopt is different
from the standard approach in the generalized statistical mechanics
context, where the modification of entropy definition leads to the
modification of algebra and consequently the derivative concept. We
adopt that the mapping to a continuous fractal space leads naturally
to the necessity of modifications in the derivatives, that we will
call deformed or metric derivatives \citep{Balankin-Towards a physics on fractals}.
The modifications of derivatives brings to a change in the algebra
involved, which in turn may conduct to a generalized statistical mechanics
with some adequate definition of entropy.

In this paper, we initiate a general variational calculus with metric
derivatives embedded into the Lagrangian. The purpose of the present
work is to develop the corresponding generalization and for this,
we define three options to pursue that will be described in the forthcoming
sections.

Our paper is outlined as follows: In Section 2, we cast some mathematical
aspects, in Section 3, we develop the variational approach based on
embedded deformed-derivatives.In Section 4, we extend the formalism
to relativistic independent fields. In Section 5, some applications
are presented; in Section 6, the Hamiltonian formalism are addressed
and we cast finally our Conclusions and Outlook in Section 7.

\section{A Glance at Mathematical Aspects}

\textbf{Hausdorff Derivative}

A model that maps hydrodynamics continuum flow in a fractal coarse-grained
(fractal porous) space, which is essentially discontinuous in the
embedding Euclidean space, into a continuous flow governed by conventional
partial differential equations was suggested in Refs. \citep{Balankin PRE85-Map-2012,Balankin Rapid Comm}.
Using non-conventional partial differential equations based on the
model of a fractal continuous flow employing local fractional differential
operators, Balankin has suggested essentially that the discontinuous
fractal flow in a fractal permeable medium can be mapped into the
fractal continuous flow, which is describable within a continuum framework,
indicating also that the geometric framework of fractal continuum
model is the three-dimensional Euclidean space with a fractal metric. 

Employing the local fractional differential operators in connection
with the Hausdorff derivative \citep{Chen-Time Fabic}, the latter
derivative can be written as \citep{Balankin PRE85-Map-2012}: 
\begin{eqnarray}
\frac{d^{H}}{dx^{\zeta}}f(x) & = & \lim_{x\rightarrow x'}\frac{f(x')-f(x)}{(x')^{\zeta}-x^{\zeta}}=\nonumber \\
 & = & \left(\frac{x}{l_{0}}+1\right)^{1-\zeta}\frac{d}{dx}f=\frac{l_{0}^{\zeta-1}}{c_{1}}\frac{d}{dx}f=\frac{d}{d^{\zeta}x}f,\label{eq:Haudorff derivative}
\end{eqnarray}
where $l_{0}$ is the lower cutoff along the Cartesian $x-$axis and
the scaling exponent, $\zeta$ $,$ characterizes the density of states
along the direction of the normal to the intersection of the fractal
continuum with the plane, as defined in the work \citep{Balankin PRE85-Map-2012}.

\textbf{Conformable Derivative}

Recently, a promising new definition of local deformed derivative,
called conformable fractional derivative, has been proposed by the
authors in Ref. \citep{new definition} that preserve classical properties
and is given by

\begin{equation}
T_{\alpha}f(t)=\lim_{\epsilon\rightarrow0}\frac{f(t+\epsilon t^{1-\alpha})-f(t)}{\epsilon}.\label{basic}
\end{equation}

If the function is differentiable in a classical sense, the definition
above yields 
\begin{equation}
T_{\alpha}f(t)=t^{1-\alpha}\frac{df(t)}{dt}.\label{eq:Differentiable}
\end{equation}

Changing the variable $t\rightarrow1+\frac{x}{l_{0}}$, we should
write (\ref{eq:Differentiable}) as $l_{0}\left(1+\frac{x}{l_{0}}\right)^{1-\alpha}\frac{d}{dx}f$,
that is nothing but the Hausdorff derivative up to a constant and
valid for differentiable functions.

Also, the Riemann Improper Integral \citep{new definition} can be
written as:

\begin{equation}
I_{a}^{\alpha}(f)(t)=\int_{a}^{t}f(x)d^{\alpha}x=\intop_{a}^{t}f(x)x^{\alpha-1}dx,
\end{equation}
where $dx^{\alpha}=\frac{1}{x^{1-\alpha}}dx.$

Another similar definition of local deformed derivative with classical
properties is that used in Ref. \citep{Aplica Katugampola}:

Let $f:[0,\infty)\rightarrow\mathbb{R}$ and $t>0$. Then the local
deformed derivative-Katugampola- of $f$ of order $\alpha$ is defined
by, 
\begin{equation}
\mathcal{D}^{\alpha}(f)(t)=\lim_{\epsilon\rightarrow0}\frac{f(te^{\epsilon t^{-\alpha}})-f(t)}{\epsilon},\label{df}
\end{equation}
 for $t>0,\;\alpha\in(0,1)$. If $f$ is $\alpha-$differentiable
in some $(0,a),\;a>0$, and $\lim_{t\rightarrow0^{+}}\mathcal{D}^{\alpha}(f)(t)$
exists, then define
\begin{equation}
\mathcal{D}^{\alpha}(f)(0)=\lim_{t\rightarrow0^{+}}\mathcal{D}^{\alpha}(f)(t).
\end{equation}

The Riemann Improper Integral is also considered in this approach.

\textbf{q-derivative in the nonextensive context}

Over the recent decades, diverse formalisms have emerged that are
adopted to approach complex systems. Amongst those, we may quote the
q-calculus in Tsallis\textquoteright{} version of Non-Extensive Statistics
with its undeniable success whenever applied to a wide class of different
systems; Kaniadakis\textquoteright{} approach, based on the compatibility
between relativity and thermodynamics; Fractional Calculus (FC), that
deals with the dynamics of anomalous transport and other natural phenomena,
and also some local versions of FC that claim to be able to study
fractal and multifractal spaces and to describe dynamics in these
spaces by means of fractional differential equations. See the references
in \citep{Nosso On conection2015}.

With the generalized nonaddictive $q-$entropy as motivation, the
$q-$derivative sets up a deformed algebra and takes into account
that the $q-$exponential is eigenfunction of $D_{(q)}$ \citep{Borges 2004}.
Borges proposed the operator for $q$-derivative as below:

\begin{equation}
{\displaystyle D_{(q)}f(x)\equiv{\displaystyle \lim_{y\to x}\frac{f(x)-f(y)}{x\ominus_{q}y}}}={\displaystyle [1+(1-q)x]\frac{df(x)}{dx}.}\label{eq:q-derivative}
\end{equation}

Here, $\ominus_{q}$ is the deformed difference operator, $x\ominus_{q}y\equiv\frac{x-y}{1+(1-q)y}\qquad(y\ne1/(q-1)).$

The q- Integral has the similar structure of Riemann Improper Integral:

\begin{equation}
\int_{a}^{t}f(x)d_{q}x=\intop_{a}^{t}\frac{f(x)}{1+(1-q)x}dx=\intop_{a}^{t}f(x)d_{q}x;
\end{equation}
where $d_{q}x=\underset{y\rightarrow x}{lim}x\ominus_{q}y=\frac{1}{1+(1-q)x}dx$.

Recently we have shown that 

\begin{equation}
1-q=\frac{(1-\zeta)}{l_{0}}.
\end{equation}

So, we concluded that the deformed $q-$derivative is the first order
expansion of the Hausdorff derivative and that there is a strong connection
between these formalism by means of a fractal metric.

Now, we highlight some results and relevant properties, which are
valid for local forms of derivative addressed here:

\textbf{Leibniz rule}

As regards the product rule for derivatives, conformable, Katugampola
or Hausdorff follows the rule similar to the usual derivatives: $D^{\alpha}(fg)=gD^{\alpha}f+fD^{\alpha}g,$
and similarly, for $q-$derivative:$D_{q}(fg)=gD_{q}f+fD_{q}g$.

I\textbf{ntegration by parts}

Generalizing, let (A,B) be a subinterval of (a, b). Consider a functional

\begin{equation}
\int_{\alpha}f(x)d^{\alpha}x=\intop_{a}^{b}f(x)x^{\alpha-1}dx.
\end{equation}

Then, one can state that the integration by parts holds.

\begin{equation}
\int_{\alpha}f(x)D^{\alpha}g(x)d^{\alpha}x=f(x)g(x)\mid_{a}^{b}-\int_{q}g(x)D^{\alpha}f(x)d^{\alpha}x.
\end{equation}

The same holds for $q-$integral.

For details, the reader can consult the refs. \citep{new definition,Aplica Katugampola}.

\textbf{The chain rule}

For composition of functions holds:

\begin{equation}
D^{\alpha}[f\circ g](x)=\frac{df(g(x))}{dg}D^{\alpha}g(x),
\end{equation}

\begin{equation}
D_{q}[f\circ g](x)=\frac{df(g(x))}{dg}D_{q}g(x).
\end{equation}

\section{Variational Approach with embedded derivatives}

Our problem here is to search minimizers of a variational problem
with deformed derivative embedded into the Lagrangian function $L.$
After the mapping into the fractal continuum, $L$ will be a C$^{2}$-function
with respect to all its arguments.

Remarks: (i)We consider a fractional variational problem which involves
local deformed-derivatives, as Hausdorff, conformable, Katugampola
or q-derivative.

(ii) The problem can be easily generalized for Lagrangians which will
depend also on higher order deformed-derivatives.

(iii) We assume that $0<\alpha<1$.

(iv) To pursue our objectives, we cast three options to address as
different approaches:

Option 1: $\alpha-$integral or $q-$integral, $\delta$usual, deformed-derivatives
embedded

Option 2: $\alpha-$integral or $q-$integral, $\delta$deformed,
deformed-derivatives embedded and

Option 3: Usual integral, $\delta$usual, deformed-derivatives embedded,
similar to Ref. \citep{Atanackovic-Variational}, but here with local
deformed or metric derivatives.

We know that deformed-kernels used here can be replaced with other
kernels, resulting in a general variational calculus, as in Ref. \citep{Agrawal}.

Our interpretation of the action above in option1 and 2 is based on
the existence of a succession of specific internal temporal intervals
for the system which lead to a anomalous dynamics \citep{Podlubny}.
The time interval can be thought as distributed with a polynomial
distributions of weights or probabilities that are convoluted with
the Lagrangian \citep{Tenreiro}. This leads for the action to be
similar to have some similarities with Stieltjes like integral. The
action then can be interpreted, in this context as an integration
of the Lagrangian, with embedded deformed-derivatives, in a cosmological
inhomogeneous time scale. It is similar to time-clock randomization
of momenta and coordinates taken from the conventional phase space
\citep{Stanislavsky} but without redefining the phase space with
some thing like fractional derivative, generalized coordinates or
momenta.

Notice that this kind of action is suitable to treat systems with
dissipative forces or non-holonomic systems since it include the scale
in time letting to consider the effects of internal times of the systems
\citep{Nabulsi}. Thus, systems with memory effects may suitably be
analyzed by means of this action. Note yet that any kind of deformed
integrals or derivatives needs to be necessarily embedded into the
Lagrangian. The only deformed integral could be the own action with
its time probability factor convoluted.

Now, we can analyze each case:

\textbf{For the option1:}

To derive the extended version of the Euler\textendash Lagrange equation
let us introduce the following $\alpha-$deformed functional action

\[
J[y]=\int_{\alpha}L(x,y,D_{x}^{\alpha}y)d^{\alpha}x,
\]
where $L$ is the Lagrangian with embedded deformed derivative $D_{x}^{\alpha}y$
and $dx^{\alpha}=\frac{1}{x^{1-\alpha}}dx$. 

Analogously, with a $q-$deformed functional we can write: 
\[
J[y]=\int_{q}L(x,y,D_{q}y)d_{q}x.
\]

We will find the condition that $J[y]$ has a local minimum. To do
so, we consider the new fractional functional depending on the parameter
$\varepsilon$ .

Consider for the variable $y(x):$

$y(x)=y^{*}(x)+\varepsilon\eta(x);$ $y^{*}(x)$ is the objective
function, and $\eta(a)=\eta(b)=0$, $\varepsilon$ is a the parameter.

So, applying the deformed derivative, we obtain:

$D_{x}^{\alpha}y(x)=D_{x}^{\alpha}y^{*}(x)+\varepsilon D_{x}^{\alpha}\eta(x).$

Using the chain rule and the well known $\delta-$variational processes
relative to the $\varepsilon$ parameter, we can write

\[
\delta_{\varepsilon}L=\frac{\partial L}{\partial y}\eta+\frac{\partial L}{\partial D_{x}^{\alpha}y}D_{x}^{\alpha}\eta.
\]

Since integration by part holds with deformed integral similarly of
integer and, using that the usual transversality condition for an
extreme value, one obtain that$\delta_{\varepsilon}J=0$ implies that
\begin{equation}
\frac{\partial L}{\partial y}-D_{x}^{\alpha}(\frac{\partial L}{\partial D_{x}^{\alpha}y})=0.
\end{equation}

If $L$depends on different derivative orders, as $L=L(x,y,D_{x}^{\alpha}y,D_{x}^{\beta}y,D_{x}^{\gamma}z),$
then one gets that the $E-L$equations can be read as

\[
\frac{\partial L}{\partial y}-D_{x}^{\alpha}(\frac{\partial L}{\partial D^{\alpha}y})-D_{x}^{\beta}(\frac{\partial L}{\partial D^{\beta}y})-(\frac{\partial L}{\partial D^{\beta}y})(\alpha-\beta)x^{-\beta}=0
\]
for $y$ variable and for $z-$variable as

\[
\frac{\partial L}{\partial z}-D_{x}^{\gamma}(\frac{\partial L}{\partial D^{\gamma}z})-(\frac{\partial L}{\partial D^{\gamma}z})(\alpha-\gamma)x^{-\gamma}=0.
\]

We can now go to the next option for analyzes.

\textbf{For the option 2:}

Here, for $\delta-$deformed approach we mean that the total derivative
now takes into account the deformed derivatives. So, we can write
for $\delta-$deformed process applied to the Lagrangian

\[
\delta_{\varepsilon}^{\alpha}L=\frac{\partial L}{\partial y}D_{\varepsilon}^{\alpha}y+\frac{\partial L}{\partial D_{x}^{\alpha}y}\frac{\partial^{\alpha}D_{x}^{\alpha}y}{\partial\varepsilon^{\alpha}},
\]
where $D_{\varepsilon}^{\alpha}y=\eta(x)\varepsilon^{1-\alpha},$
$\frac{\partial^{\alpha}D_{x}^{\alpha}y}{\partial\varepsilon^{\alpha}}=D_{x}^{\alpha}\eta(x)\varepsilon^{1-\alpha}.$
Now, using the integration by parts, one gets

\begin{equation}
\frac{\partial L}{\partial y}-D_{x}^{\alpha}(\frac{\partial L}{\partial D_{x}^{\alpha}y})=0.
\end{equation}

For $q-$derivative we can similarly write:

\begin{equation}
\frac{\partial L}{\partial y}-D_{q,x}(\frac{\partial L}{\partial D_{q,x}y})=0.
\end{equation}

We can see that the results are analogous to those obtained by the
option 1 approach.

We can pursue now results from the next option.

\textbf{For the option 3:}

The Lagrangian $L$ can be $L=L(x,y,D_{x}^{\alpha}y,D_{x}^{\beta}y,D_{x}^{\gamma}z),$
where embedded deformed derivatives are considered again.

In this case we consider the fractional action $J[y]=\int L(x,y,D_{x}^{\alpha}y,D_{x}^{\beta}y,D_{x}^{\gamma}z)dx$
and the usual $\delta-$variational processes

\[
\delta_{\varepsilon}L=\frac{\partial L}{\partial y}\eta+\frac{\partial L}{\partial D_{x}^{\alpha}y}D_{x}^{\alpha}\eta_{1}+\frac{\partial L}{\partial D_{x}^{\beta}y}D_{x}^{\beta}\eta_{1}+\frac{\partial L}{\partial z}\eta_{2}+\frac{\partial L}{\partial D_{x}^{\gamma}z}D_{x}^{\gamma}\eta_{2},
\]
with $D_{\varepsilon}^{\alpha}\eta=x^{1-\alpha}\frac{d\eta}{dx}$
and so on.

The resulting $E-L$ equations are:

\begin{equation}
\frac{\partial L}{\partial y}-\frac{d}{dx}(x^{1-\alpha}\frac{\partial L}{\partial D_{x}^{\alpha}y})-\frac{d}{dx}(x^{1-\beta}\frac{\partial L}{\partial D_{x}^{\beta}y})=0,
\end{equation}

\begin{equation}
\frac{\partial L}{\partial z}-\frac{d}{dx}(x^{1-\gamma}\frac{\partial L}{\partial D_{x}^{\gamma}z})=0.
\end{equation}

For $q-$derivative we will have:

\begin{equation}
\frac{\partial L}{\partial y}-l_{01}\frac{d}{dx}[(1+(1-q_{1}x))\frac{\partial L}{\partial D_{q_{1},x}y}]-\frac{d}{dx}[(1+(1-q_{2}x))\frac{\partial L}{\partial D_{q_{2,}x}y}]=0,
\end{equation}

\begin{equation}
\frac{\partial L}{\partial z}-l_{02}\frac{d}{dx}[(1+(1-q_{3}x))\frac{\partial L}{\partial D_{q_{3},x}z}]=0.
\end{equation}

\section{Relativistic, independent Fields}

Now, we can proceed to pursue equivalent approaches to field theory,
based on independent relativistic fields.

Here, $\phi=\widetilde{\phi}+\epsilon_{1}^{\mu}\delta\phi,$ $\psi=\widetilde{\psi}+\epsilon_{2}^{\mu}\delta\psi,$ 

\[
\partial_{\mu}\phi=\partial_{\mu}\widetilde{\phi}+\epsilon_{1}^{\mu}\partial_{\mu}\delta\phi,
\]

\[
\partial_{\mu}\psi=\partial_{\mu}\widetilde{\psi}+\epsilon_{2}^{\mu}\partial_{\mu}\delta\psi.
\]

Here,$\delta\phi,\delta\psi$ are arbitrary,$\widetilde{\phi},$$\widetilde{\psi}$are
the objective fields and $\mu=0,1,2,3,$ following the index spatial-temporal
derivative, $\partial_{\mu}.$

With non-deformed standard derivative, we usually consider the fractional
action

\[
S=\int dt\int d^{3}x\mathcal{L}(\phi,\partial_{\mu}\phi,\psi,\partial_{\mu}\psi,x^{\mu}),
\]
with the usual $\delta-$ process

\[
\delta_{\epsilon}\mathcal{L=\frac{\partial\mathcal{L}}{\partial\phi}\delta\phi}+\frac{\partial\mathcal{L}}{\partial\partial_{\mu}\phi}\delta\partial_{\mu}\phi+\frac{\partial\mathcal{L}}{\partial,\psi}\delta\psi+\frac{\partial\mathcal{L}}{\partial\partial_{\mu}\psi}\delta\partial_{\mu}\psi.
\]

For the deformed derivative embedded into the Lagrangian we can set

$\partial_{\mu}\rightarrow\partial_{\mu}^{\alpha_{\lambda}}$, $\lambda=0,1,2,3;$
$\mathcal{\mathcal{L=}L}(\phi,\partial_{\mu}^{\alpha_{\lambda}}\phi,\psi,\partial_{\mu}^{\beta_{\lambda}}\psi,x^{\mu})$.

For Option 1,2, the deformed Euler-Lagrange Equations can be written
as

\begin{equation}
\frac{\partial\mathcal{L}}{\partial\phi}-\partial_{\mu}^{\alpha_{\lambda}}(\frac{\partial\mathcal{L}}{\partial(\partial_{\mu}^{\alpha_{\lambda}}\phi)})=0
\end{equation}
 and analogously for $\psi$.

For the option 3 approach, we have: 
\begin{equation}
\frac{\partial\mathcal{L}}{\partial\phi}-\partial_{\mu}[(x^{\mu})^{1-\alpha_{\lambda}}\frac{\partial\mathcal{L}}{\partial(\partial_{\mu}^{\alpha_{\lambda}}\phi)}]=0.
\end{equation}

The possible inclusion of higher order derivatives could lead to an
interesting study of Podolski-like systems and will be let to a future
publication.

Now we can cast some possible applications to the approaches here
stated.

\section{Applications:}

\subsection{Deformed Newtonian Mechanics}

We consider in this example the ``free quasi-particle'', with $V=0$
and $L=\frac{1}{2}m(D_{t}^{\alpha}x)^{2}$

For the options 1,2, we have similar results as in the Ref. \citep{Fractional Newton mechanics with conformable}

For option3, we obtain interesting results that are cast below.

Te $E-L$equation is 

\[
\frac{\partial L}{\partial x}-\frac{d}{dt}(t^{1-\alpha}\frac{\partial L}{\partial D_{t}^{\alpha}x})=0,
\]
 that leads to the movement equation

\begin{equation}
2(1-\alpha).m.t^{1-2\alpha}\frac{dx}{dt}+t^{2-2\alpha}.m\frac{d^{2}x}{dt^{2}}=0.
\end{equation}

Note that we can do some variable transformation, so that $t\equiv1+\frac{t'}{l_{0}}$
$\Longrightarrow$ $dt=\frac{dt'}{l_{0}}$. This means that here we
are dealing with the Hausdorff derivative and there is no problems
in the point $t'=0.$ The dynamical equation can now be written as

\begin{equation}
2(1-\alpha).m.(1+\frac{t'}{l_{0}})^{1-2\alpha}\frac{dx}{dt'}+(1+\frac{t'}{l_{0}})^{2-2\alpha}.m\frac{d^{2}x}{dt'^{2}}=0
\end{equation}

For a low level fractionality, we define $(1-\alpha)\equiv\epsilon.$
With first order in $\epsilon,$ it can be written as 
\begin{equation}
2\epsilon m\frac{dx}{dt'}+(1+2\epsilon.(1+\frac{t'}{l_{0}}).ln(1+\frac{t'}{l_{0}}))m\frac{d^{2}x}{dt'^{2}}=0,
\end{equation}
The equation stated above is in some way a time dependent-mass equation
with the presence of friction, showing the appearance of dissipation
due to the complex context. This is an speculated result, since we
are considering an open system.

Here we would like to observe certain resemblances with some results
in Ref. \citep{Fractional Newton mechanics with conformable}, but
in that paper it seems that the focus was not the variational approach
but only some attempt to construct a deformed Newtonian mechanics,
the derivative treated there was only the conformable one and the
result on eq. (34) (Ref. \citep{Fractional Newton mechanics with conformable})
seems to has an absent $t-$factor in the dissipative term, as the
reader can verify.

Analogously, for $q-$derivative in the nonextensive statistics context
we can obtain the $E-L$equations as

\[
\frac{\partial L}{\partial x}-\frac{d}{dt}[l_{0}(1+(1-q)t)\frac{\partial L}{\partial D_{q,t}x}]=0,
\]

that leads to the dynamical movement equations as

\[
2l_{0}^{2}(1-q)m[1+(1-q)t]\frac{dx}{dt}+l_{0}^{2}[1+(1-q)t]^{2}m\frac{d^{2}x}{dt^{2}}=0.
\]

For low level fractionality, we again define $1-q\equiv\epsilon,$
so, for the first order in $\epsilon$ parameter results that:

\[
2\epsilon m\frac{dx}{dt}+(1+2\epsilon t)m\frac{d^{2}x}{dt^{2}}=0.
\]
 The equations above contains a time-dependent mass and there is again
the presence of friction. For $V\neq0$ and $\epsilon\rightarrow0,$
$l_{0}\rightarrow1,$ the dissipative and the time-dependent term
vanishes and we re-obtain the classical Newtonian law for conservative
forces:

\[
m\frac{d^{2}x}{dt^{2}}=-\frac{\partial V}{\partial x}.
\]
So, the results are evidently consistent with the classical Newtonian
mechanics.

\subsection{Deformed Time-Deformed-Schr\"odinger Equation}

We now proceed with the constructions related to a deformed- field
theory, specially that leads to some mass-dependent Schr\"odinger equations.

To Pursue this objective, let us now consider the Lagrangian density

\[
\mathcal{L=}i\hbar\psi^{*}D_{t}^{\alpha}\psi+\frac{\hbar^{2}}{2m}\nabla\psi^{*}.\nabla\psi-V\psi^{*}\psi
\]
and take into account the options 1,2.

The $E-L$ equation leads to the deformed-time-dependent Schr\"odinger
equation (note that $\nabla$implies to $\alpha=1$ order in the spatial
part of the deformed Euler-Lagrange equations)

\begin{equation}
i\hbar D_{t}^{\alpha}\psi=-\frac{\hbar^{2}}{2m}\nabla^{2}\psi-V\psi.
\end{equation}

In terms of q-derivative, analogously we can show that the resulting
$E-L$ equation is

\begin{equation}
i\hbar D_{q,t}\psi=-\frac{\hbar^{2}}{2m}\nabla^{2}\psi-V\psi=H\psi,\label{eq:q-Schrodinger}
\end{equation}

Here, let us redefine the $q-$derivative as a new $scale-q-$derivative
\begin{equation}
{\displaystyle D_{(q)}^{\lambda}f(\lambda x)\equiv}{\displaystyle [1+(1-q)\lambda x]\frac{df(x)}{dx}.}\label{eq:q-derivative redef}
\end{equation}

The solution of eq.(\ref{eq:q-Schrodinger}) with this $scale-q-$deformed-derivative
is 

$\psi=Ae_{q}(-\frac{i}{\hbar}Ht)$.

We can observe in additions that a solution of the equation of the
nonlinear Schr\"odinger equation $i\hbar\partial_{t}\psi=H\psi^{q},$
for $V=0,$ is also $\psi=Ae_{q}(-\frac{i}{\hbar}Ht)$ and we can
write that

\[
i\hbar\frac{\partial}{\partial t}\psi=-\frac{\hbar^{2}}{2m}\nabla^{2}\psi^{q}.
\]
That is, the nonlinear Schr\"odinger equation called in Refs. \citep{NRT-PRL-2011}
(with $q=q'-2$ compared to the $q-$index of the reference) as NRT-like
Schr\"odinger equation and can be thought as resulting from a $time-scale-q-$deformed-derivative
applied to the wave function $\psi$.

\textbf{Deformed Spatial Schr\"odinger equation}

Let us now consider the Lagrangian density expressed by means of the
spatial deformed derivatives as

\begin{equation}
\mathcal{L=}i\hbar\psi^{*}\overset{.}{\psi}+\frac{\hbar^{2}}{2m}D_{x}^{\alpha}\psi^{*}.D_{x}^{\alpha}\psi-V\psi^{*}\psi.
\end{equation}

With option 1,2 the resulting dynamical equation, that is, the spatial
deformed-Schr\"odinger equation is

\begin{equation}
i\hbar\frac{\partial}{\partial t}\psi=-\frac{\hbar^{2}}{2m}D_{x}^{\alpha}D_{x}^{\alpha}\psi+V\psi.
\end{equation}

In terms of $q-$derivative, we can show that a similar result is

\begin{equation}
i\hbar\frac{\partial}{\partial t}\psi=-\frac{\hbar^{2}}{2m}D_{q,x}D_{q,x}\psi+V\psi.
\end{equation}
 Both results are possible expressions of the position-dependent mass,
similarly as CAFA-like Schr\"odinger equation in Ref. \citep{Costa-Almeida-}.

\subsection{Option 3: Deformed Schr\"odinger Equation}

Consider again the Lagrangian density as

$\mathcal{L=}i\hbar\psi^{*}\overset{.}{\psi}+\frac{\hbar^{2}}{2m}D_{x}^{\alpha}\psi^{*}.D_{x}^{\alpha}\psi-V\psi^{*}\psi.$

By the approach in the option 3, we obtain a form of Cauchy-Euler
for Schr\"odinger equation

\[
i\hbar\frac{\partial}{\partial t}\psi=-\frac{\hbar^{2}}{2m}[2(1-\alpha)x^{-\alpha}D_{x}^{\alpha}\psi+x^{2-\alpha}\frac{\partial^{2}\psi}{\partial x^{2}}]+V\psi.
\]

For \textbf{$q-$}derivative we obtain

\begin{equation}
i\hbar\frac{\partial}{\partial t}\psi=-\frac{\hbar^{2}}{2m}[2(1-q)l_{0}^{2}(1+(1-q)x)\frac{\partial\psi}{\partial x}+(1+(1-q)x)^{2}l_{0}^{2}\frac{\partial^{2}\psi}{\partial x^{2}}]+V\psi.
\end{equation}

For low fractionalily we again define $1-q=\epsilon$ and considering
only first order in $\epsilon$we can write a position-dependent mas
Schr\"odinger equations as
\begin{equation}
\hbar\frac{\partial}{\partial t}\psi=-\frac{\hbar^{2}}{2m}[2(\epsilon l_{0}^{2}\frac{\partial\psi}{\partial x}+(1+2\epsilon x)l_{0}^{2}\frac{\partial^{2}\psi}{\partial x^{2}}]+V\psi.
\end{equation}

\section{Hamiltonian Formalism}

We will now pursue the Hamiltonian formalism that takes into account
the embedded deformed derivatives proposed.

\textbf{For the option 1:} 

Consider $L=L(t,q,D_{t}^{\alpha}q)$, where $q$ is some generalized
coordinate and have not to be confused with the entropic parameter. 

The total derivative of $L$ is written as

\begin{equation}
dL=\frac{\partial L}{\partial t}dt+\frac{\partial L}{\partial q_{i}}dq_{i}+\frac{\partial L}{\partial(D_{t}^{\alpha}q_{i})}d(D_{t}^{\alpha}q_{i}).
\end{equation}

By the $E-L$ equation, we can obtain that: $\frac{\partial L}{\partial q_{i}}=D_{t}^{\alpha}\frac{\partial L}{\partial(D_{t}^{\alpha}q_{i})}.$
Defining now the generalized momentum as $p_{i}^{\alpha}=\frac{\partial L}{\partial(D_{t}^{\alpha}q_{i})},$
we obtain

\begin{equation}
dL=\frac{\partial L}{\partial t}dt+D_{t}^{\alpha}\frac{\partial L}{\partial(D_{t}^{\alpha}q_{i})}dq_{i}+p_{i}^{\alpha}d(D_{t}^{\alpha}q_{i}).
\end{equation}

Following, defining $H=H(p_{i}^{\alpha},q_{i},t)\equiv p_{i}^{\alpha}(D_{t}^{\alpha}q_{i})-L$,
we will have

\begin{eqnarray}
dH & = & \frac{\partial H}{\partial t}dt+\frac{\partial H}{\partial q_{i}}dq_{i}+\frac{\partial H}{\partial p_{i}^{\alpha}}dp_{i}^{\alpha}=dp_{i}^{\alpha}D_{t}^{\alpha}q_{i}+p_{i}^{\alpha}d(D_{t}^{\alpha}q_{i})-dL\nonumber \\
 & = & dp_{i}^{\alpha}D_{t}^{\alpha}q_{i}+p_{i}^{\alpha}d(D_{t}^{\alpha}q_{i})-\frac{\partial L}{\partial t}dt-D_{t}^{\alpha}p_{i}^{\alpha}dq_{i}+\frac{\partial L}{\partial(D_{t}^{\alpha}q_{i})}d(D_{t}^{\alpha}q_{i})-p_{i}^{\alpha}d(D_{t}^{\alpha}q_{i}).
\end{eqnarray}
Thus, we obtain

\begin{equation}
dH=dp_{i}^{\alpha}D_{t}^{\alpha}q_{i}-D_{t}^{\alpha}p_{i}^{\alpha}dq_{i}-\frac{\partial L}{\partial t}dt.
\end{equation}

Comparing the the equations above we can state that

\begin{equation}
\begin{cases}
\frac{\partial H}{\partial t}= & -\frac{\partial L}{\partial t},\\
\frac{\partial H}{\partial q_{i}}= & -D_{t}^{\alpha}p_{i}^{\alpha},\\
\frac{\partial H}{\partial p_{i}^{\alpha}}= & D_{t}^{\alpha}q_{i}.
\end{cases}
\end{equation}

\textbf{For the option 2}:

We can rewrite: $dL\rightarrow d^{\alpha}L$

\begin{equation}
d^{\alpha}L=\frac{\partial L}{\partial t}d^{\alpha}t+\frac{\partial L}{\partial q_{i}}d^{\alpha}q_{i}+\frac{\partial L}{\partial(D_{t}^{\alpha}q_{i})}d^{\alpha}(D_{t}^{\alpha}q_{i})
\end{equation}
or

\begin{eqnarray}
\frac{d^{\alpha}L}{dt^{\alpha}} & = & \frac{\partial L}{\partial t}D_{t}^{\alpha}t+D_{t}^{\alpha}(\frac{\partial L}{\partial(D_{t}^{\alpha}q_{i})})D_{t}^{\alpha}q_{i}+\frac{\partial L}{\partial(D_{t}^{\alpha}q_{i})}D_{t}^{\alpha}(D_{t}^{\alpha}q_{i}).\nonumber \\
 & = & D_{t}^{\alpha}[p_{i}^{\alpha}D_{t}^{\alpha}q_{i}]+\frac{\partial L}{\partial t}D_{t}^{\alpha}t,
\end{eqnarray}
thus, we can write

$D_{t}^{\alpha}[p_{i}^{\alpha}D_{t}^{\alpha}q_{i}-L]=-\frac{\partial L}{\partial t}D_{t}^{\alpha}t.$ 

Defining again $H$ as $H\equiv p_{i}^{\alpha}D_{t}^{\alpha}q_{i}-L=H(p_{i}^{\alpha},q_{i},t)$
and differentiating results that

\begin{eqnarray*}
\frac{d^{\alpha}H}{dt^{\alpha}} & = & \frac{\partial H}{\partial t}D_{t}^{\alpha}t+\frac{\partial H}{\partial q_{i}}(D_{t}^{\alpha}q_{i})+\frac{\partial H}{\partial p_{i}^{\alpha}}(D_{t}^{\alpha}p_{i}^{\alpha})\\
 & = & (D_{t}^{\alpha}p_{i}^{\alpha}).(D_{t}^{\alpha}q_{i})+(p_{i}^{\alpha}).D_{t}^{\alpha}(D_{t}^{\alpha}q_{i})-D_{t}^{\alpha}L.\\
 & = & (D_{t}^{\alpha}p_{i}^{\alpha}).(D_{t}^{\alpha}q_{i})-\frac{\partial L}{\partial t}D_{t}^{\alpha}t-D_{t}^{\alpha}(\frac{\partial L}{\partial(D_{t}^{\alpha}q_{i})})D_{t}^{\alpha}q_{i}.
\end{eqnarray*}

$\frac{\partial H}{\partial t}D_{t}^{\alpha}t+\frac{\partial H}{\partial q_{i}}(D_{t}^{\alpha}q_{i})+\frac{\partial H}{\partial p_{i}^{\alpha}}(D_{t}^{\alpha}p_{i}^{\alpha})=-\frac{\partial L}{\partial t}D_{t}^{\alpha}t-D_{t}^{\alpha}(\frac{\partial L}{\partial(D_{t}^{\alpha}q_{i})})(D_{t}^{\alpha}q_{i})+(D_{t}^{\alpha}q_{i}).(D_{t}^{\alpha}p_{i}^{\alpha})$

Comparing again the equations, results that

\begin{equation}
\begin{cases}
\frac{\partial H}{\partial t}=- & \frac{\partial L}{\partial t}\\
\frac{\partial H}{\partial q_{i}}=- & D_{t}^{\alpha}(\frac{\partial L}{\partial(D_{t}^{\alpha}q_{i})})=-D_{t}^{\alpha}(p_{i}^{\alpha})\\
\frac{\partial H}{\partial p_{i}^{\alpha}}= & D_{t}^{\alpha}q_{i}
\end{cases}
\end{equation}

\textbf{For the option 3:} 

We can show that the resulting equations are (remembering that the
$E-L$ in this case is given by $\frac{\partial L}{\partial q_{i}}-\frac{d}{dt}(t^{1-\alpha}\frac{\partial L}{\partial D_{t}^{\alpha}q_{i}})=0$)

\[
\begin{cases}
\frac{\partial H}{\partial t}=- & \frac{\partial L}{\partial t}\\
\frac{\partial H}{\partial q_{i}}=- & \frac{d}{dt}(t^{1-\alpha}p_{i}^{\alpha})=(\alpha-1)t^{-\alpha}p_{i}^{\alpha}-D_{t}^{\alpha}(p_{i}^{\alpha})\\
\frac{\partial H}{\partial p_{i}^{\alpha}}= & D_{t}^{\alpha}q_{i}.
\end{cases}
\]

To finish this section we want to outline some possible extension
for the formalism developed here.

One of such possibilities is to extend to the Pontryagin's maximum
(or minimum) principle used in optimal control theory.

This formalism can be used in the presence of constraints for the
state or input controls and will yield a set of deformed-differential
equations to study open systems that are interacting with the environment.

\subsection{Noether procedure with deformed derivative}

Let us now consider the Lagrangian density as $\mathcal{L}(\phi_{i},\partial_{\mu}\phi_{i}).$With
the usual $\delta-$ process, we obtain for the variation of $\mathcal{L}:$

\[
\delta\mathcal{L}=\frac{\partial\mathcal{L}}{\partial\phi_{i}}\delta\phi_{i}+\frac{\partial\mathcal{L}}{\partial\partial_{\mu}\phi_{i}}\delta\partial_{\mu}\phi_{i}.
\]

With $\mathcal{L}(\phi_{i},\partial_{\mu}^{\alpha_{\mu}}\phi_{i}),$
we shall have

\[
\delta\mathcal{L}=\frac{\partial\mathcal{L}}{\partial\phi_{i}}\delta\phi_{i}+\frac{\partial\mathcal{L}}{\partial(\partial_{\mu}^{\alpha_{\mu}}\phi_{i})}\delta(\partial_{\mu}^{\alpha_{\mu}}\phi_{i}).
\]

Integrating by parts, considering the $E-L$ equations and assuming
that $\delta\mathcal{L}$ is given by a total deformed-derivative
of order $\alpha_{\mu}$ of some admissible function $G^{\mu},$ $\delta\mathcal{L}=(\partial_{\mu}^{\alpha_{\mu}}G^{\mu}),$
we write

\[
\partial_{\mu}^{\alpha_{\mu}}[\frac{\partial\mathcal{L}}{\partial(\partial_{\mu}^{\alpha_{\mu}}\phi_{i})}(\delta\phi_{i})-G^{\mu}]=0.
\]

The conserved generalized current can be defined as

\[
J_{\alpha_{\mu}}^{\mu}=\frac{\partial\mathcal{L}}{\partial(\partial_{\mu}^{\alpha_{\mu}}\phi_{i})}(\delta\phi_{i})-G^{\mu}.
\]

Explicit specific cases, like gauge currents, energy-momentum, angular
momentum, spin tensors and their corresponding invariances shall be
discussed elsewhere in a forthcoming work.

\section{Conclusions and Outlook}

In conclusion, we have employed the variational calculus to obtain
$E-L$ equations with deformed derivatives.

The paradigm that governs the standard approach in the context of
generalized statistical mechanics was revisited.

To achieve our goals, Lagrangian and Hamiltonian formalisms were studied,
some physical reasoning about integration based on the existence of
succession of specific internal temporal intervals for the system
are furnished and a new class of generalized Lagrangian, Hamiltonian,
and action principles are presented.

Position-dependent mass equations and nonlinear equations could be
shown to result from the formalism that are in agreement with results
found in the scientific literature.

Possible extensions of the subject and the concepts discussed here
are also outlined, such as Pontryagin's maximum (or minimum) principle.

We believe that with this formalism can set up a systematic way to
obtain nonstandard equations in several areas of science, without
an excessive heuristics. This can avoid to introduce \textit{ad hoc}
fields and unnecessary suppositions about strange dynamics.

As an outlook, one of our studies regards the Landau-Lifshitz-Gilbert-Slonczewski
equations in the context os metric derivatives and complex systems,
to analyse the effects on the damping.

....

The authors wish to express their gratitude to FAPERJ-Rio de Janeiro
and CNPq-Brazil for the partial financial support.

We thank Prof. Bruno G. Costa and Prof. Ernesto P. Borges for very
valuable discussions.\textbf{\bigskip{}
}

\end{document}